# NUCLEAR CORRECTIONS FOR CROSS SECTION OF LEPTON INELASTIC SCATTERING

*Dmitry Timashkov*[*]

Moscow Engineering Physics Institute, Moscow 115409, Russia
*Experimental Complex NEVOD*

**Abstract**

Main nuclear corrections for inelastic scattering of charged leptons on nuclei, namely shadowing, EMC-effect and Fermi motion have been investigated. Simple formulas describing these effects for different $x$-regions have been proposed, and unified expression for any transferred energy and momentum as well as for different nuclei has been obtained. Results of calculations of the ratio of nucleus structure function (per nucleon) and free nucleon structure function are in a good agreement with SLAC data.

**1. Introduction**

Inelastic scattering on nuclei plays important role in muon propagation at large depths of rock or water. In comparison with other electromagnetic processes, inelastic cross section more slowly decreases with the increase of transferred momentum and relative inelastic energy loss logarithmically rises with lepton energy. However to describe inelastic energy loss in matter it is necessary along with proton structure functions to take into account nuclear corrections.

There are three main nuclear effects in inelastic scattering: shadowing, EMC-effect, and Fermi motion. Shadowing is the decrease of structure functions of bound nucleon in comparison with free nucleon structure function. It happens when hadronic excitations (or fluctuations) produced by a real or virtual high-energy photon propagate over distances (in the laboratory frame) which are comparable to or larger than the distance between neighboring nucleons, and coherently scatter on them. Two others corrections have incoherent nature. EMC-effect appears as excess over the unity of the ratio of bound nucleon structure function to the free one at $0.1 \leq x \leq 0.2$ (anti-shadowing) and subsequent decrease of this ratio with the increase of Bjorken variable $x$ in the region $0.2 < x < 0.6$. Such behavior is caused by distortion of virtual pion field due to nuclear surrounding. The third nuclear correction is connected with Fermi motion of nucleon in nucleus and leads to a significant enhancement of the ratio for large values of Bjorken variable ($x \sim 1$).

---

[*] timashkov@nevod.mephi.ru

All these effects are clearly seen in various experiments where inelastic electron (muon) scattering on different nuclei are studied (see [1–4]). For each effect various theoretical descriptions were proposed (see reviews [5, 6]). In this paper a unified simple formula taking into account all nuclear effects in inelastic scattering of charged leptons on nuclei is developed.

## 2. Main definitions

In charged lepton scattering on unpolarized nuclear targets the differential cross section is defined by two structure functions in a similar manner as for free nucleon. For discussion of the data of various experiments it is convenient to use structure functions which depend on the Bjorken scaling variable for a free nucleon[†], $x = Q^2/(2M\nu)$:

$$\frac{d^2\sigma^A}{dxdQ^2} = \frac{4\pi\alpha_e^2}{Q^4} A \left[ \left(1 - y - \frac{Mxy}{Q^2}\right) \frac{F_2^A(x,Q^2)}{x} + y^2 F_1^A(x,Q^2) \right]. \tag{1}$$

Here $F_2^A$ is structure function of nucleus divided by atomic mass number $A$. One can say that $F_2^A$ is structure function of a bound nucleon. Usually, the ratio of $F_2^A$ for various nuclei to structure function of nucleons bound in deuterium $F_2^d$ is investigated:

$$r_A = \frac{F_2^A}{F_2^d}. \tag{2}$$

Neglecting nuclear effects in deuterium, the function $F_2^d$ can be considered as nucleon structure function averaged on isospin $F_2^N = (F_2^p + F_2^n)/2$. If nuclear effects are absent the value of $r_A$ is equal to the unity.

Experimental data [1] evidence that the value $R$ (ratio of longitudinal and transverse polarized virtual photon cross sections which determines the relation between $F_1$ and $F_2$) does not depend on atomic mass. Therefore differential cross section for inelastic scattering on nuclei can be written in the following form:

$$\frac{d^2\sigma_A}{dxdQ^2} = A r_A(x,Q^2) \frac{d^2\sigma_N}{dxdQ^2}. \tag{3}$$

Generally, the value $r_A$ depends on both kinematic variables: $Q^2$ and $x$.

Above-mentioned nuclear effects have different nature and result in the deviations of $r_A$ from the unity in different regions of $x$. Therefore, to describe the influence of nuclear

---

[†] Here $M$ is nucleon mass, $Q^2$ and $\nu$ are transferred 4-momentum and energy, correspondingly. The variable $y = \nu / E$, where $E$ is lepton energy.

corrections on proton structure function, three main effects will be considered independently, and

$$r_A = r_{sh} \cdot r_{EMC} \cdot r_F, \tag{4}$$

where $r_{sh}$ describes shadowing, $r_{EMC}$ takes into account EMC-effect and $r_F$ reflects the influence of Fermi-motion of nucleon in nucleus.

## 3. Three nuclear effects

### 3.1. Shadowing

The shadowing reveals itself as a reduction of cross section for scattering of real photon on nucleus with respect to sum of cross sections for scattering on nucleons forming the nucleus. A similar effect is observed for virtual photon at low values of $x$:

$$\frac{d^2\sigma_A}{dxdQ^2} < A\frac{d^2\sigma_N}{dxdQ^2}, \tag{5}$$

or

$$F_2^A < F_2^N. \tag{6}$$

In various approximations of this phenomenon [1, 7–8] the power-like form is used:

$$F_2^A \sim A^{-\beta} F_2^N, \tag{7}$$

where β is small. More accurately experimental data [9] for photon inelastic scattering on nuclei can be described by means of the following dependence [10]:

$$F_2^A \approx \left(0.22 + 0.78 A^{-0.11}\right) F_2^N. \tag{8}$$

This form implies that with probability 22 % the photon is a point-like object and in other 78 % interacts through hadron fluctuations.

In lepton inelastic scattering the shadowing is maximal at $x = 0$ and vanishes at $x \geq 0.1$. To describe $r_{sh}$-dependence on $x$, the following function is proposed:

$$r_{sh} = \left(0.22 + 0.78 A^{-0.11}\right)\exp(-x/x_\pi) + \left(1 - \exp(-x/x_\pi)\right), \tag{9}$$

which leads to dependence (8) at $x = 0$ and is close to 1 at $x \geq 0.1$. The parameter $x_\pi$ defines typical value of Bjorken variable, at which the correction caused by the changing of pion field becomes large (see the next subsection). The typical value of $x_\pi$ is close to $m_\pi/M \simeq 0.15$ [5].

### 3.2. EMC-effect

EMC-effect is observed as enhancement of $F_2^A$ above $F_2^N$ by several percent in region $0.1 < x < 0.2$ and subsequent decrease of $r_A$ till $x \sim 0.6$ where Fermi-motion begins to play a significant role. There are several models explaining such behavior of $r_A$: $x$-rescaling, $Q^2$- rescaling, cluster models, and so on. The simplest way to take into account EMC-effect consists in the use of linear approximation of $r_A$ in this region (see for example [11]):

$$r_{EMC} \approx 1 + a_{EMC} - b_{EMC} x, \tag{10}$$

where as a first approximation the parameters $a_{EMC}$ and $b_{EMC}$ do not depend on atomic mass number.

To eliminate the influence of EMC-effect in the region $x \to 0$ which has been already described by expression (9), it is necessary to put $r_{EMC}(x=0) = 1$. For this purpose, a formula similar to (9) is proposed:

$$r_{EMC} = 1 + a_{EMC}\left(1 - \exp(-x/x_\pi)\right) - b_{EMC} x. \tag{11}$$

In a broad range of elements ($A > 10$) parameters $a_{EMC}$ and $b_{EMC}$ can been taken as:

$$a_{EMC} = 1.3, \quad b_{EMC} = 0.7. \tag{12}$$

*3.3. Fermi-motion*

Fermi-motion of nucleons in nucleus significantly changes the $x$-dependence of nucleus structure function. It happens due to differences in kinematics of inelastic scattering on rest nucleon and on nucleon which flies with momentum of the order of Fermi momentum $p_F$. These differences appear as enhancement of $F_2^A$ with respect to $F_2^N$ in the region $x \sim 1$. A standard way of calculation of the correction on Fermi-motion is a convolution of structure function of free nucleon with momentum distribution function of nucleons in nucleus [5–6]:

$$F_2^A = \int_x^A dz\, f_N(z)\, F_2^N(x/z), \tag{13}$$

where $f_N(z)$ is the momentum distribution function for nucleon, which has $z$ fraction of nuclear momentum (in Breit frame).

For ideal Fermi-gas approximation, $f_N(z)$ can be written in the analytical form:

$$f_N(z) = \frac{3z}{4\xi_F}\left(1 - \frac{(1-z)^2}{\xi_F^2}\right) \cdot \theta\left(1 - \frac{|1-z|}{\xi_F}\right), \tag{14}$$

where $\xi_F = p_F/M$, $\theta$ is Heaviside step function. The value of $p_F$ is about 270 MeV.

In the region $x \sim 1$, nucleon structure function has a form:

$$F_2 \sim x_P (1 - x_P)^{3 + 4t/3}, \tag{15}$$

which follows from solution of the evolution equations in the limit $x \to 1$ (see [12]). Here $t$ is a evolution variable:

$$t = \frac{12}{33 - 2n_f} \ln \frac{\ln\left((Q^2 + Q_0^2)/\Lambda^2\right)}{\ln\left(Q_0^2/\Lambda^2\right)}, \tag{16}$$

$$n_f = 3, \quad Q_0^2 = 4 \text{ GeV}^2, \quad \Lambda = 230 \text{ MeV},$$

and $x_P$ is a unified scaling variable [13]:

$$x_P = \frac{2x}{1 + \sqrt{1 + 4M^2 x^{1+x}/Q^2}}. \tag{17}$$

In this case the expression for $r_F$ has the following form:

$$r_F(x, Q^2) = \int_x^A \frac{x_P(x/z)}{x_P(x)} \frac{(1 - x_P(x/z))^{3+4t/3}}{(1 - x_P(x))^{3+4t/3}} f_N(z) dz. \tag{18}$$

It should be emphasized that, in contrast to other nuclear corrections, the value $r_F$ may strongly depend on $Q^2$. Indeed, since at $x \to 1$ the nucleon structure function tends to zero and, though $F_2$ depends on $Q^2$ in a logarithmical way (see (15)–(16)), even small changes of exponent can induce significant deviations of $r_F$ (Fig. 1).

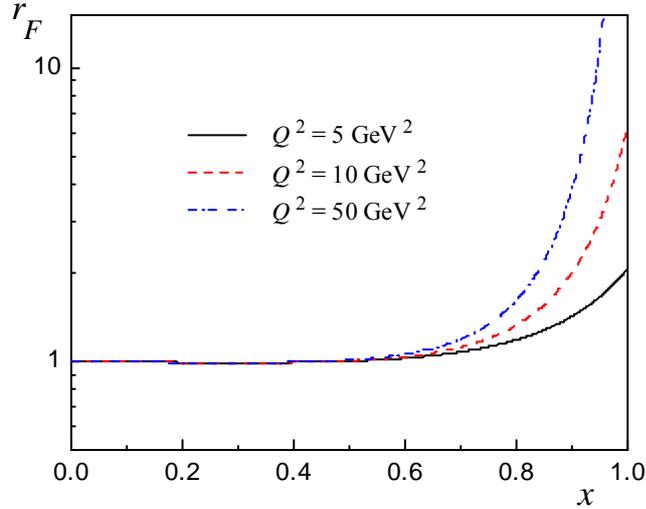

Fig. 1. Ratio of structure functions $F_2^A$ and $F_2^N$ taking into account only Fermi motion.

## 4. Results and discussion

Using expressions (9), (11), (18) and also (4), one can calculate the dependences of $r_{sh}$, $r_{EMC}$, $r_F$ and correspondingly of $r_A$. Results of calculations for carbon, aluminum, iron and silver are presented in Fig. 2. Experimental data were averaged over the acceptance range of $Q^2$ from 2 GeV$^2$ till 15 GeV$^2$ (see [1]). However for $x > 0.6$ experimental points mainly correspond to $Q^2 = 10$ GeV$^2$ therefore the calculations have been performed with this value of $Q^2$.

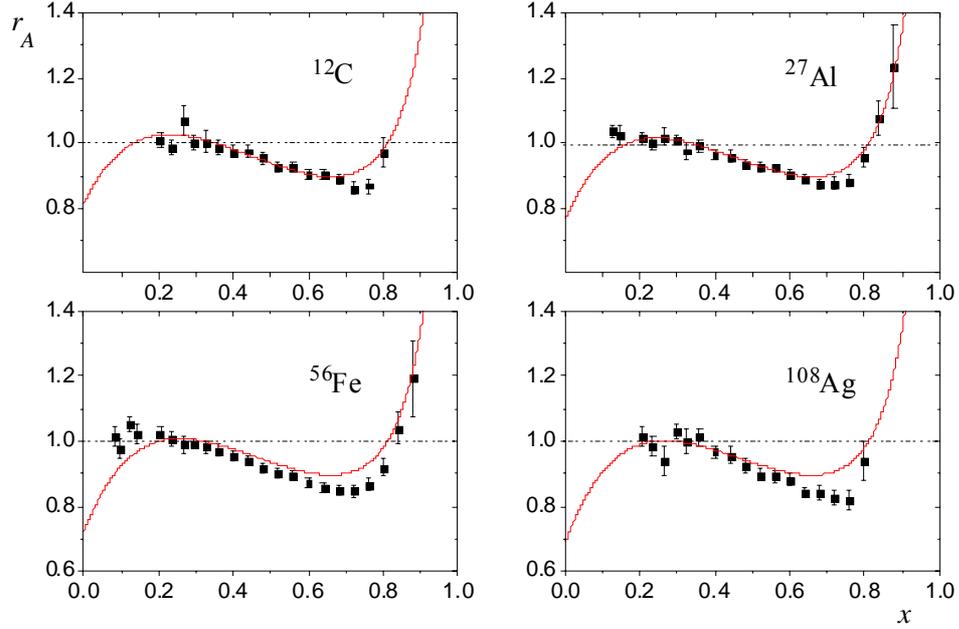

*Fig. 2. Dependences of ratio of the bound nucleon structure function to the free nucleon structure function. Experimental data are from [1].*

A good agreement with experimental data for light nuclei confirms that the proposed method of description of nuclear effects is sufficiently effective even with the simplest choice of dependences describing nuclear corrections. These results are useful for calculations of total inelastic scattering charged lepton cross section and for estimations of lepton inelastic energy loss in rock and water. Some deviations for heavy nuclei can be apparently explained by a slight logarithmic dependence of $a_{EMC}$ and $b_{EMC}$ on atomic mass number.

**5. Acknowledgements**

Author would like to thanks Prof. A. A. Petrukhin for initialization of this work, fruitful discussions and useful advices.